\documentclass[a4paper]{jpconf}
\usepackage{graphicx}
\begin{document}
\title{Physics to plan AWAKE Run~2}

\author{Patric Muggli for the AWAKE Collaboration}

\address{Max Planck Institute for Physics, F{\"o}hringer Ring 6, Munich 80805, Germany}

\ead{muggli@mpp.mpg.de}

\begin{abstract}
We briefly describe the basic physics principles considered for planning of the AWAKE Run~2 experiment. %
These principles are based on experimental results obtained during Run~1 and knowledge obtained from numerical simulation results and other experiments. %
The goal of Run~2 is to accelerate an electron bunch with a narrow relative energy spread and an emittance sufficiently low for applications. %
The experiment will use two plasmas, electron bunch seeding for the SM process, on-axis external injection of an electron bunch and electron bunch parameters to reach plasma blow-out, beam loading and beam matching. %
\end{abstract}

\section{Introduction}
Relativistic proton bunches produced by the CERN SPS (400\,GeV) or LHC (6.5\,TeV) carry large amounts of energy (19 and 104\,kJ, respectively) and are therefore potential drivers of wakefields over long plasma length. %
Externally injected electrons could then be accelerated and reach the 100s of GeV to TeV energy level. %
These bunches are long ($\sim$(6-12)\,cm) and thus require self-modulation~\cite{bib:kumar} (SM) to effectively drive large amplitude wakefields ($\sim$1\,GV/m). %

AWAKE~\cite{bib:muggli}, the Advanced WAKefield Experiment aims to explore the possibility of using these proton bunches for plasma wakefield acceleration of electrons. %
It also aims to explore applications of this scheme to high-energy physics~\cite{bib:hep}. %
AWAKE Run~1 results very successfully met the goal that were set forward: demonstration of the SM of the SPS proton bunch in the 10\,m plasma~\cite{bib:marlene,bib:karl} and acceleration of externally injected, 18\,MeV electrons to the GeV energy level~\cite{bib:nature}. %
One of the major experimental results is that the proton bunch SM is reproducible when seeded at sufficient level~\cite{bib:fabian}. %

The plans for Run~2 are based on the results of Run1 and on the goal to accelerate a particle bunch with a narrow relative energy spread and an emittance sufficiently low for applications high-energy physics applications ((10-20)\,mm-mrad). %

\section{Run~1 results implications for Run~2}

Run~1 yielded a wealth of other results that those mentioned above, and much was learned about the SM and the injection process. %

Numerical simulations with AWAKE parameters show that SM drives wakefields that, while non-linear (i.e., with amplitude a significant fraction of the wave breaking fields)~\cite{bib:pukhov}, they do no preserve the emittance of an externally injected electron bunch. %
Therefore, electron bunch parameters must be chosen to preserve emittance. %

Simulations show that in a uniform plasma the accelerating field decreases after saturation of the seeded SM process~\cite{bib:pukhov,bib:caldwell}. %
These simulations~\cite{bib:caldwell} also show that, with a plasma up-density step (a few \%) located in the early stage of the SM development ($\sim$(1-2)\,m after the plasma entrance), the wakefields retain at large fraction of their saturation amplitude after the SM process. %

Experiments performed in Run~1 used a relativistic ionization front for seeding of the SM process. %
This seeding method leaves the front of the bunch, ahead of the ionization front, not modulated since it propagates in rubidium (Rb) vapor and does not interact with it. %
Experiments have also shown that SM develops in a preformed plasma as SM instability (SMI), i.e., without any seeding and all along the bunch~\cite{bib:gessner}. %
Self-modulation instability of the front of the bunch in a following, preformed plasma would generate wakefields that would interfere with those driven by the self-modulated back of the bunch. %
This can be avoided for example by seeding the SM process with a preceding electron bunch. %

The plasma source has density ramps at the entrance and exit~\cite{bib:genady}. %
In the entrance ramp, the transverse wakefields driven by the unmodulated bunch are mostly focusing for  protons and therefore defocusing for electrons. %
In addition, the phase velocity of the wakefields is slower than that of the proton bunch and non constant till the location of saturation of the SM process along the plasma~\cite{bib:pukhov}. %
Wakefields therefore dephase with respect to the externally injected, relativistic electrons that may be defocused and lost. %
To avoid these effects, external injection of electrons must occur after the SM process has saturated and on axis to preserve emittance. %
Run~1 thus used injection at an angle with respect to, and some distance into the plasma~\cite{bib:nature}. %

These arguments then lead to the plan for Run~2 experiments. %
These experiments should show that much higher energies (tens to hundreds of GeV) could in principle be reached "simply" by making the accelerator plasma longer. %
We therefore also develop other plasma sources to produce long plasma lengths~\cite{bib:sources} satisfying the AWAKE stringent density requirements: less than 0.25\% density variations over 10\,m. %

We describe here the principles that we plan on implementing for the next experiments. %

\subsection{Run~2 experimental layout}

\begin{figure}[h]
\includegraphics[width=35pc]{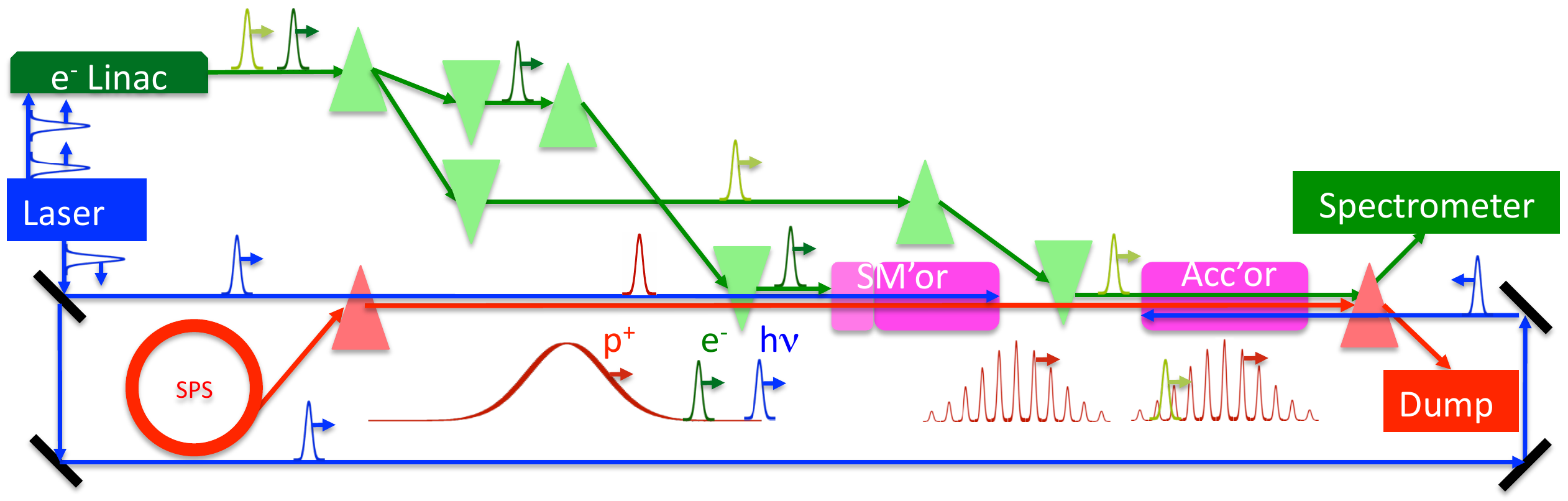}
\caption{\label{fig:run2layout}Layout for Run~2 experiments. Experimental parameters in Table~\ref{table:expparams}.}
\end{figure}
Figure~\ref{fig:run2layout} shows a schematic layout of the experiment for Run~2. %
Table~\ref{table:expparams} gives the general experimental parameters. %
\begin{table}[]
\centering
\caption{Beam, vapor, plasma and laser parameters of the AWAKE Run 2 experiment. The first plasma column has the same density step as the first Rb column and there is the same gap between plasmas as between Rb columns.}
\label{table:expparams}
\begin{tabular}{@{}lllll}
\hline
\textbf{Parameter}                         & \textbf{Symbol}   & \textbf{Value}         & \textbf{Range}                & \textbf{Unit} \\ \hline
\textit{$p^+$ Bunch}     		   &                             &                               &                                         &                     \\ \hline
Energy                                          & W$_{0p}$                  & 400                        &-                                         & GeV          \\ 
Relativistic Factor                         & $\gamma_{0p}$         & 427                        &-                                         &-          \\ 
Population                                    & N$_p$                          & $3\times10^{11}$    &     $(1-3)\times10^{11}$            & $p^+$ /bunch \\ 
Length                                          & $\sigma_{zp} $        & 12                        &6-12                                       & cm \\ 
Focused Size                               & $\sigma_{r0p} $        & 200                        &-                                       & $\mu$m \\ 
Normalized Emittance                  & $\epsilon_{Np} $       & $3.5\times10^{-6}$  &-                                      & m-rad \\ 
$\beta$-function at Waist             & $\beta_{0p}=\frac{\gamma_{0p}\sigma_{r0p}^2}{\epsilon_{Np}}$         & 5                              &-                                       & m \\ 
Relative Energy Spread               & $\Delta W_{0p}/W_{0p}$  & 0.03\%                   &-                                       &-  \\ \hline
\textit{Rubidium Vapor}               &              &                                  &                                        &                     \\ \hline
Density (Both)                                      & n$_{Rb}$            & $7\times10^{14}$   &   $(1-10)\times10^{14}$  & cm$^{-3}$          \\ 
Column Length (Both)                 & L$_{Rb}$            & 10                           &-                                       & m          \\ 
Columns Radius (Both)                   & r$_{Rb}$             & 2                            &-                                       & cm          \\ 
Density Step (1$^{st}$ Column)   & $\Delta$n$_{Rb}$/n$_{Rb}$      &-                    &   (0-10)\%                     &-        \\ 
Density Step Location                & L$_{s}$                     & -                    &   0.5, 1.0, ..., 4.0                     & m        \\ 
Gap Between Columns                  & g             & 0.3                            &-                                       & m          \\ \hline
\textit{Fiber/Ti:Sapphire Laser}                                  &  \textit{}&                               &                                        &                    \\ \hline
Central Wavelength                       & $\lambda_{0}$       & 780                       &-                                        & nm          \\ 
Bandwidth                                     & $\Delta\lambda_{0}$ & $\pm$5                   &-                                        & nm          \\ 
Pulse Length                                & $\tau_{0}$              & 120                       &-                                        & fs          \\ 
Max. Compressed Energy    & $E_{max}$              & 450                       &$\le$200 per Rb column                                        & mJ          \\ 
Focused Size                              & r$_{l}$                    & 1                            &-                                        & mm          \\ 
Rayleigh Length                           & Z$_{r}$                    & 4                       &-                                        & m          \\ \hline
\textit{Plasma}                              &                                &                               &                                         &                    \\ \hline
Electron Density (Both)              & n$_{e0}$            & $7\times10^{14}$   &   $(1-10)\times10^{14}$  & cm$^{-3}$          \\ 
Electron Plasma Frequency                       & f$_{pe}$          & 237                      &90-284                          & GHz          \\ 
Electron Plasma Wavelength                      & $\lambda_{pe}$ & 1.3                  &3.3-1.1                                       & mm          \\ 
Length (Both)                                  & L$_{p}$            & 10                           &-                                       & m          \\ 
Radius (Both)                                     & r$_{p}$             & $>$1                            &-                                       & mm          \\ \hline
\textit{e$^-$ Bunch for Acceleration~\cite{bib:veronica}}                              &                                &                               &                                         &                    \\ \hline
Energy                                          & W$_{0e}$                  & 218                        &-                                         & MeV          \\ 
Relativistic Factor                         & $\gamma_{0e}$         & 427(=$\gamma_{0p}$)                        &-                                         &-          \\ 
Population                                    & N$_e$                          & $6.2\times10^{8}$    &     $(3-30)\times10^{8}$            & $e^-$ /bunch \\ 
Length                                          & $\sigma_{ze} $        & 60                        & -                                    & $\mu$m \\ 
Normalized Emittance                  & $\epsilon_{Ne} $       & $2\times10^{-6}$  &-                                      & m-rad \\ 
Matched $\beta$-function at Waist & $\beta_{0e}$         & 5.9                              &-                                       & mm \\ 
Matched Focused Size                 & $\sigma_{r0e} $        & 5.25                        &-                                       & $\mu$m \\ 
Beam to Plasma Density Ratio   &n$_{be}$/n$_{e0}$         & 35                            & -                                     & $\mu$m \\ 
Relative Energy Spread               & $\Delta W_{0e}/W_{0e}$  & 0.1\%                   &-                                       &-  \\ \hline
\end{tabular}
\end{table}
The plasma is split into a self-modulator and an accelerator, both $\sim$10\,m-long, separated by a $\sim$30\,cm gap allowing for on-axis injection of the electron bunch. %
The self-modulator plasma includes a density step. %
Injection then occurs in the accelerator plasma driven by the fully self-modulated proton bunch~\cite{bib:livio}. %
After the plasmas, the proton bunch travels to a dump while the electron bunch travels to an energy spectrometer and other diagnostics (not shown). %

The seeding method must lead to SM of the entire proton bunch. %
Seeding is thus provided by a short electron bunch preceding the proton bunch. %
Electron bunch parameters need to be determined for optimum seeding of the SM process~\cite{bib:mugglieseed}. %
Since a Rb vapor is used here, the electron bunch is itself preceded by the ionization laser pulse that pre-forms the plasma (see Fig.~\ref{fig:run2layout}). %
Wakefields driven by a self-modulated bunch do no preserve emittance of the accelerated electron bunch. %
The witness bunch parameters must be adjusted for reaching electron blow-out for the majority of the bunch charge in order to preserve slice emittance. %
Additional constraint comes from the necessity of loading the wakefields to minimize the relative energy spread and contribute to preservation of the projected emittance. %
Imagining a very long plasma corresponding to many betatron oscillations of the electrons and thus possibly of the bunch envelope size, it is desirable to match the bunch to the ion column focusing force to maintain beam matching and avoid large variation in the electron bunch divergence at the plasma exit. %

Seeding SM with an electron bunch replaces the relativistic ionization front seeding method used in Run~1.  %
Therefore, two electron bunches must be produced either by two separate linacs, or by the same linac (as shown schematically on Fig.~\ref{fig:run2layout}).
The parameters of the two electron bunches may be quite different, and we are currently studying the best method to produce them. %
The seed electron bunch must only provide sufficient wakefields amplitude ($\sim$MV/m) over the first few meters of plasma. %
Low energy may thus be sufficient (e.g., $\sim$20\,MeV). %

Toy numerical simulations were performed to have a first estimate for the parameters of the electron bunch to be accelerated~\cite{bib:veronica}. %
Parameters presented here are extracted from this study that used as input: normalized emittance $\epsilon_{Ne}$=2\,mm-mrad, relativistic factor $\gamma_{0e}$=427, plasma electron density n$_{e0}$=7$\times$10$^{14}$\,cm$^{-3}$. %
They must be recalculated for self-consistent parameter sets compatible with Run~2 plans and capabilities. %

The witness bunch must be short to occupy a small fraction of the wakefields period ($\lambda_{pe}$=1.3\,mm) and effectively load the wakefields. %
To reach blow-out within its length, its density must exceed n$_{be min}\cong\frac{\lambda_{pe}}{\sigma_z}n_{e0}$=20$n_{e0}$. %
In order to match the electron bunch at the plasma entrance it must be focused to a very tight transverse size. %
The matching condition to the ion column of density n$_{i0}$=n$_{e0}$ reads:  $\frac{\sigma_{r0e}^4\gamma_{0e}n_{e0}}{\epsilon_{Ne}^2}=\frac{2\epsilon_0m_ec^2}{e^2}$. %

\subsection{Plasma sources}

The plasma sources (self-modulator and accelerator) are initially of the same type as for Run~1: Rb vapor ionized by a short, intense laser pulse~\cite{bib:oz}. %
This choice follows from the fact that this source is the only one known to produce the plasma density uniformity AWAKE requires. %
Moreover, it is in principle straightforward to implement the step in the Rb vapor density of the self-modulator. 

A temperature step is imposed on the Rb vapor of the self-modulator. %
This temperature steps translates into a Rb density step, which is then turned into a plasma density step by laser ionization. %
Numerical simulation results~\cite{bib:genady} show that for a step-function in temperature, the Rb (an thus plasma) density has a continuous, $\sim$10\,cm-long step. %
This ramp is sufficiently short to preserve the effect on the developing wakefields~\cite{bib:lotovpriv}. %

This scheme will also allow for comparison between the relativistic ionization seeding of Run~1 and the electron beam seeding of Run~2. %

The accelerator source is ionized by a second laser pulse traveling against the self-modulated proton and injected electron bunches. %
In this scheme, the relative timing is such that the electron bunch and the ionizing laser pulse meet each other slightly downstream from the entrance of the accelerator source. %
The electron bunch thus enters a plasma with ``infinitely'' sharp boundary, though relativistically moving against it. %
It therefore avoids any possible plasma density ramp effects, as well as any effects created by the dense plasma generated by the ionizing laser pulse on a vacuum window or laser pulse dump foil. %
Initial simulations show that this process does not change the electron bunch incoming parameters~\cite{bib:petrenko}. %

The self-modulator source is $\sim$10\,m long, enough to guarantee that the SM process saturates before exiting the source at the two plasma densities chosen for most experiments. %
These two densities are 2 and 7$\times$10$^{14}$\,cm$^{-3}$. %
The low density is that at which the streak camera time resolution ($\sim$1\,ps) is sufficient to directly visualize the effect of the SM process on the proton bunch (see~\cite{bib:karl}). %
The high density is the highest at which acceleration experiments were performed and at which the largest energy gain was observed (see~\cite{bib:nature}). %
The source will consist of heating segments whose temperature can be individually controlled to vary the position and height of the density step. %
This should allow for verification and optimization of the effect. %
The source will include view ports for Rb density measurements, as well as for shadowgraphy to measure the plasma density perturbation associated with the wakefields driving. %
Simulation results indicate that a change in density perturbation supporting wakefields by a factor of four can be expected with and without the density step. %
 
The accelerator source is also $\sim$10\,m long, enough for the injected electron bunch ($\sim$165\,MeV) to reach multi-GeV energies. %
It is essentially a simplified and improved version of the Run~1 source. %
This is because the plasma density uniformity is most important along that source to maintain proper relative phasing between the wakefields and the accelerating electrons. %

The source is followed by an electron energy spectrometer and by optical diagnostics for the proton bunch (as in Run~1). %
Relative timing diagnostics are also developed for the synchronization of the proton, electron bunches and laser pulses at the injection point. %

As mentioned above, an accelerator plasma source whose length can be made from tens of meters to hundreds or thousands of meters with suitable density uniformity for acceleration is required for HEP applications. %
The length of laser-ionized, alkali metal vapor sources is limited by depletion of the laser pulse energy, to a few tens of meters. %
We are developing other possible sources such as discharge and helicon sources~\cite{bib:sources}. %
The helicon source has a unit cell structure consisting of a RF and a magnetic coil. %
Such cells can be stacked to reach very long plasma lengths. %
However, both source types still have to demonstrate sufficient density uniformity. %
 
\subsection{Electron bunch for acceleration}

\begin{figure}[h]
\begin{minipage}{18pc}
\includegraphics[width=18pc]{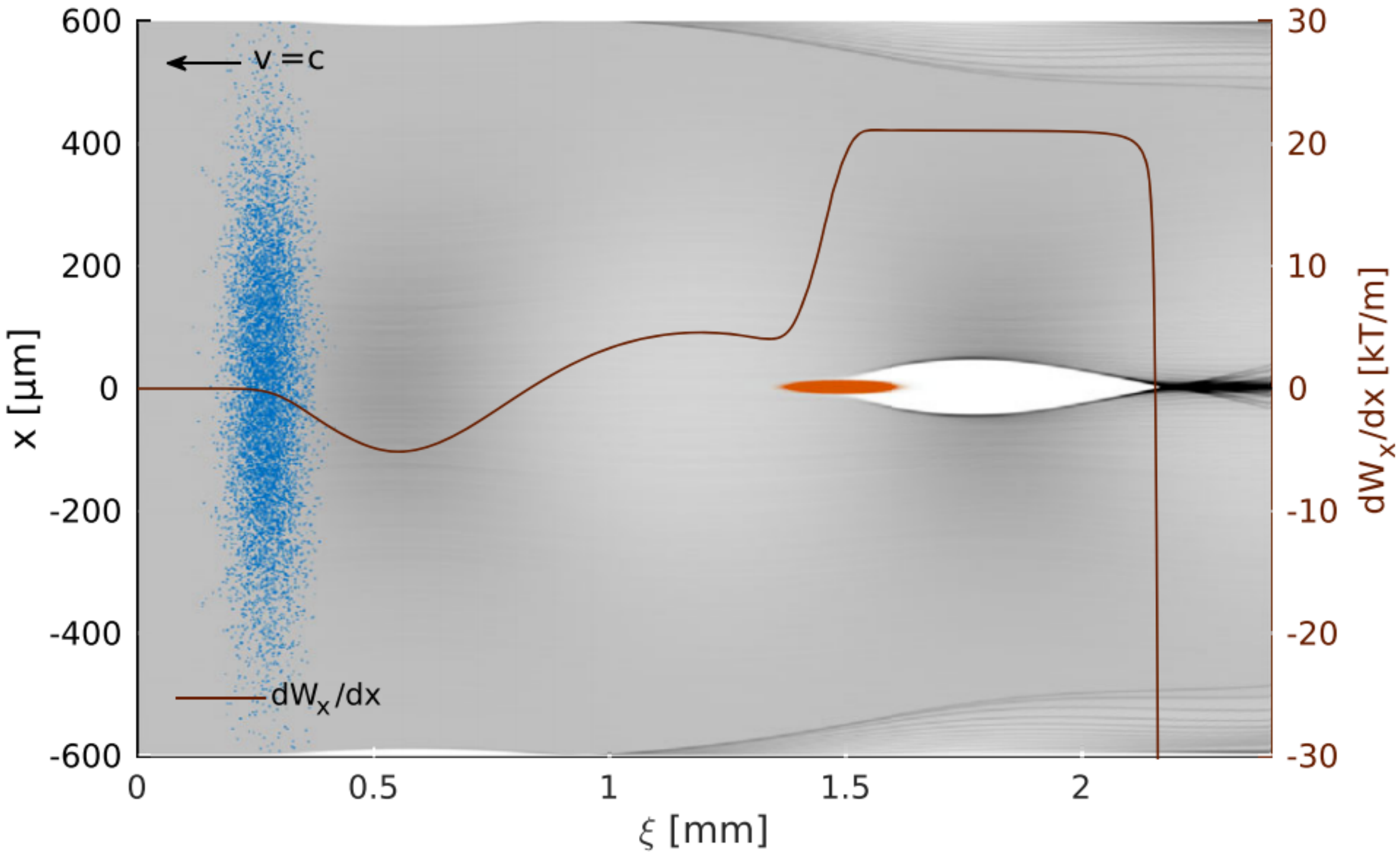}
\caption{\label{fig:runebeam}Numerical simulation result showing in grey the plasma electron density in a slice along ($\xi$) and across (x) the simulation window. %
The single, short drive proton bunch is in blue and electron bunch in red, both moving to the left. %
Blow-out of plasma electrons (white region) is reached with the electron bunch. %
The red line shows the focusing field gradient (kT/m), evaluated at the propagation axis. %
It is maximum and constant along $\xi$ in the blow-out region shown by the white area ($1.6\le\xi\le1.8$\,mm, white = zero plasma electron density). %
From Ref.~\cite{bib:veronica}.}
\end{minipage}\hspace{2pc}%
\begin{minipage}{18pc}
\includegraphics[width=18pc]{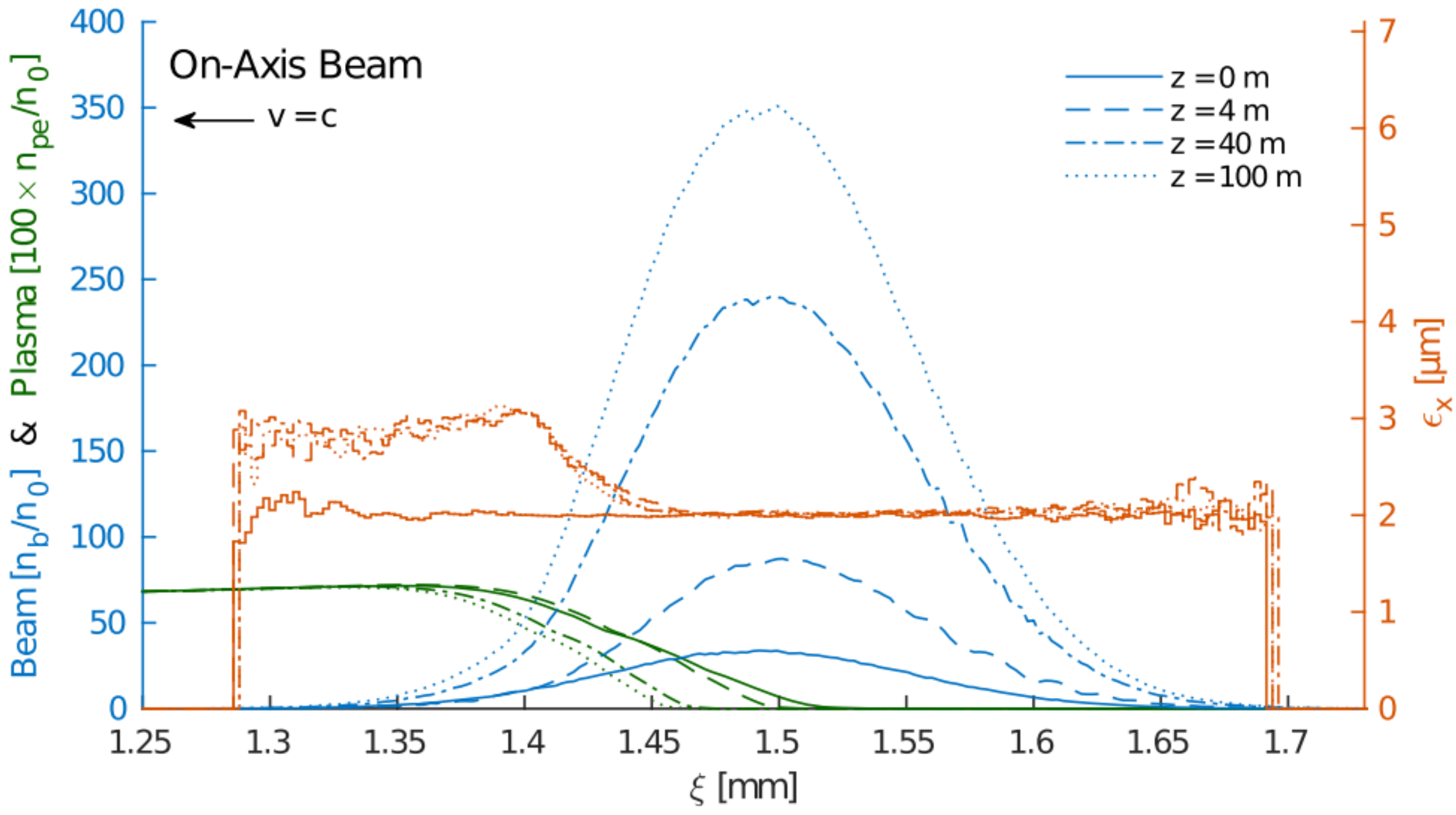}
\caption{Bunch density (blue lines) along the beam axis $\xi$ at four different propagation positions $z$ in the plasma. %
Red lines show a moving window calculation of the transverse normalized emittance at the same propagation distances. %
The plasma density profiles (green lines) are multiplied by a factor of 100 to be visible. %
These simulations were run with an LHC energy drive proton beam of 7\,TeV. %
From Ref.~\cite{bib:veronica}.} 
\label{fig:emitpreserv} 
\end{minipage} 
\end{figure}
The principle behind acceleration of a quality electron bunch is based on well-known plasma wakefield principles. %
The electron bunch parameters were determined in toy-model numerical simulations using a single, non-evolving proton bunch driving wakefields similar to those driven by the self-modulated bunch~\cite{bib:veronica} (see Fig.~\ref{fig:runebeam}). %
The purpose was to demonstrate that the majority of the electron bunch charge can be accelerated with incoming normalized emittance preservation at the 2\,mm-mrad level and small final energy spread (at the \% level) in a plasma with density ${\rm n}_{e0}=7\times10^{14}$\,cm$^{-3}$. %
In order to preserve the bunch slice emittance, the bunch density n$_{be}$ must exceed that of the plasma and for a bunch length of 60\,$\mu$m, this corresponds to ${\rm n}_{be}/{\rm n}_{e0}=35$. %
The electron bunch initial relativistic factor is equal to that of the protons ($\gamma_{0e}=\gamma_{0p}=427$). %

In order to be matched to the ion focusing force, the bunch beta-function at the waist and at the plasma entrance $\beta_{0e}$ must be 5.9\,mm. %
This corresponds to a small focused transverse size of 5.25\,$\mu$m. %
Producing such a short and tightly focused bunch requires electrons with $\sim$150\,MeV energy  to mitigate space charge effects. %
In order to load the wakefields, the charge in the bunch must be 100\,pC (6.2$\times$10$^8$\,e$^-$). %
Simulation results show that with these parameters, 73\% of the bunch charge sees its incoming emittance preserved (see Fig.~\ref{fig:emitpreserv}) and the final energy spread is smaller than 1\%. %
This is true even after acceleration over 100\,m of plasma, with the energy reaching 33.7\,GeV. %
We note that in the planned experiments the electron bunch energy may be a bit lower (150 to 165\,MeV). %
This only lightly changes the expected results since the matched beta-function scales with $\sqrt{\gamma_{0e}}$ and the matched transverse size with $\gamma_{0e}^{-1/4}$. %
Initial simulations including misalignment of the electron bunch with respect to the proton bunch axis, show that misalignment on the order of the electron bunch transverse size does not significantly affect the final bunch parameters. %

We are performing numerical simulations with the proton bunch self-modulating in the first plasma that include the density step, propagating in vacuum between plasmas, over an assumed gap length of 30\,cm, to determine the optimum electron bunch parameters at the two main operating densities. %
We are using a larger normalized emittance than in the toy model, 20\,mm-mrad, sufficient for possible HEP applications. %

The challenge in the injection scheme is to design the experimental region between the two plasma sources with sufficient diagnostics to measure the electron bunch parameters: waist location, transverse size, pointing and timing between the ionizing laser pulse (as a time reference for the wakefields) and the electron bunch. %

We note here that reaching blow-out, beam loading beam matching to the pure ion plasma focusing force are features that have been reached or tested and other experiments. %
There is thus no doubt that with suitable beam plasma parameters these features can also be reached in AWAKE. %

\subsection{Other studies}

The hose instability~\cite{bib:witthum} has often been posited as being a limitation for the acceleration length in plasma-based accelerators. %
While hosing of the drive proton bunch was not observed during Run~1 in the normal density operating range of 1 to 10$\times$10$^{14}$\,cm$^{-3}$, it was observed at very low plasma densities~\cite{bib:huether}. %
The Run~2 experimental setup will also include time resolved images of the proton bunch transverse charge distribution (see~\cite{bib:karl,bib:karl2}) in both transverse planes and possibly at two distances from the plasma source exit. %
Understanding actual noise sources for hosing, its growth and devising and testing methods to mitigate or suppress its possible appearance is key for very long accelerator lengths. %

\subsection{Summary}

After the successful completion of Run~1, we are developing plans for Run~2. %
These plans are based on lessons learned during Run~1, on basic principle of plasma wakefield acceleration and on preliminary simulation results. %
Successful acceleration of an externally injected electron bunch to the few GeV level, with finite energy spread (\% level) and preserved incoming normalized emittance (at the 10 to 20\,mm-mrad) would allow us to contemplate first applications of the acceleration scheme to high-energy physics. %

\end{document}